\documentclass[12pt]{article}
\usepackage{latexsym}
\usepackage{epsf}

\def\be{\begin{equation}}
\def\ee{\end{equation}}
\def\calF{{\cal F}}

\def\calG{{\cal G}}

\def\calM{{\cal M}}
\def\calN{{\cal N}}
\def\calA{{\cal A}}

\def\prodn{\prod_{n=1}^\infty}

\def\Bd{| B \rangle}

\def\Bdz{| B \rangle^{(0)}}
\def\Bz{|0\rangle}

\def\bidet{\det(C_{\mu\nu})}
\def\kdet{\det(C^k_{\mu\nu})}
\def\pdet{\prod_k\det(C^k_{\mu\nu})}
\def\rdet{\sqrt{\det(C_{\mu\nu})}}
\def\rkdet{\sqrt{\det(C^k_{\mu\nu})}}

\def\mm{{\cal M}_1 {\cal M}_2^T}

\newcommand{\e}{{\rm e}}

\begin{document}
\begin{flushright}
\footnotesize
\footnotesize
SWAT-236\\
{\bf hep-th/9908071}\\
August, $1999$
\normalsize
\end{flushright}

\begin{center}

\vspace{1.5cm}
{\Large {\bf Towards a Supersymmetic Classification of}}\\
\vspace{.4cm}
{\Large {\bf D--Brane Configurations with Odd Spin Structure}}

\vspace{1.5cm}


{\bf D\'{o}nal O'Driscoll}

\vspace{.4cm}

{
{\it Department of Physics, University of Wales Swansea \\
Singleton Park, Swansea, SA2 8PP, UK} \\
{\tt pydan@swansea.ac.uk}
}

\vspace{.5cm}
\vspace{.3cm}
\vspace{.8cm}


{\bf Abstract}
\end{center}

\begin{quotation}
\small

We consider the construction of a general tree level amplitude for the interactions between dynamical D--branes where the configurations have non--zero odd spin structure. Using Riemann Theta Identities we map the conditions for the preservation of some supersymmetry to a set of integer matrices satisfying a simple but non--trivial equation. We also show how the regularization of the RR zero modes plays an important role in determining which configurations are permitted.

\end{quotation}


\newpage
\section{Introduction}
One of the most important ways for probing the non-perturbative aspects of superstrings and M--theory is the use of D$p$--branes \cite{Polchinski:1996na}-\cite{Dai:1989ua}. These extended objects can be described by the string theoretic boundary state formalism \cite{Ademollo:1975pf}-\cite{dunbar} which provides BRST invariant solutions and allows many features of the branes to be expressed in a simple way. In particular it gives relative ease for turning on of electro--magnetic fields on the branes and the application of Lorentz boosts and rotations \cite{Abouelsaood:1987gd}-\cite{Billo:1997eg}. From duality, the results derived from this approach can also provide information regarding configurations of D--branes with NS--branes and M--branes.

The formalism is also useful for understanding the properties of the tree level amplitudes formed when two D--branes interact via a closed string. Supersymmetry arguments require that this amplitude will vanish as long as the configuration of branes retain at least a fraction of the supersymmetry. For the most part this has been done by utilizing the properties of the {\it abtruse identity}. However, it is also required that the contribution to the amplitude from the odd RR sector vanishes. Since this is proportional to the theta function $\theta_{1,1}(z_i)$ this occurs when the parameter $z_i$ is equal to zero.

Recently, it has become apparent that there is a variety of different configurations of D--branes where $\theta_{1,1}(z_i) \neq 0$. Moreover, these configurations preserve a fraction of the supersymmetry despite the non--zero contribution from the odd RR sector. This is the case for the generic situation when we apply all possible transformations and turn on all the fields in a configuration between two branes or a brane and an anti--brane \cite{SheikhJabbari:1997cv, Kitao:1998vn}. They also arise when one examines configurations of a brane with a brane - anti--branes system \cite{Billo:1998vr} and when the interacting branes are Hodge dual to each other \cite{Bertolini:1998mg} . It the understanding of how these cases preserve supersymmetry and what fraction thereof that this paper will investigate. 

In the process of this we examine which configurations are actually permitted. It is not trivial to consider configurations of any D$p - $D$p^\prime$ system due to complications arising from the fermionic zero modes in the RR sector. The standard normalization permits only cases were $|p - p^\prime| = 0, 8$. However, in the presence of fields and transformations the normalization is altered and the allowed values of $p - p^\prime$ changes. The resolution to handling these changes is to use a technique originally developed by Yost \cite{yost} which we will generalize to more arbitary configurations. 

In section 2 we construct the boundary state for a D$p$--brane carrying arbitary fields and under a general Lorentz transformation. In section 3 we evaluate the corresponding tree level amplitudes and study the regularization method needed to handle the RR zero modes. Section 4 deals with Riemann Theta Identities and it is shown how they can be used to classify the supersymmetry of the amplitudes in terms of integer valued matrices.

An earlier work \cite{SheikhJabbari:1997cv} made progress in this direction and offered a proof of one of the Riemann Theta Identities. We extend this work and show how their results are a specific example of ours.

\section{Boundary States}

We assume a flat D$p$-brane in ten dimensions upon which we perform various Lorentz transformations and turn on external electric and magnetic fields. The general boundary term in the action performing this takes the form
\be
S_b = \oint d\sigma C_{\mu\nu}X^{\mu} {\partial \over \partial \phi}X^{\nu}
\ee
corresponding to constant field strengths and velocities $C_{\mu\nu}$; and where $\phi = \sigma,\, \tau$ appropriately. The constraints on the boundary state $\Bd$ are given by summing over $\nu$, {\it viz}.
\be
C_{\mu\nu}{\partial \over \partial \phi} X^{\nu} \Bd = \left( {\partial \over \partial \sigma} ( \calG^{\mu\nu} X^{\nu}) +  {\partial \over \partial \tau}(\calF^{\bar{\mu}\nu} X^{\nu}) \right) \Bd = 0 \label{eq:boundary}
\ee
Imposing BRST and Virasoro invariance will project out many of the $C_{\mu\nu}$. Using stan\-dard no\-ta\-tion, we will set $\phi= \sigma$ for $\nu = 0, \ldots, p$ (Neu\-man/longitudinal directions) and $\phi = \tau$ for $\nu = p+1, \ldots, 9$ (Dirichlet/transverse conditions).

The well known solution to~(\ref{eq:boundary}) is
\be
\Bd = \calN \e^{\sum_n {1 \over n}a_n^{\dagger\mu}\calM_{\mu\nu}\tilde{a}_n^{\dagger\nu}} \Bd^{(0)} \Bd^{gh}
\ee
where $\calM_{\mu\nu} = (M\,S)_{\mu\nu} \in SO(1,9)$ is the matrix formed from the electromagnetic fields and Lorentz boosts and rotations, $C_{\mu\nu}$; and fixing the value of $p$ by choosing the $\psi$ such that
\be
S_{\mu\nu} = (\eta_{\alpha \beta}, -\delta_{ij})
\ee
where $\alpha$, $\beta$ label the Neumann directions and $i$, $j$ label Dirichlet directions. The advantage of this notation is that the action of T--duality alters $S$ by exchanging signs, but leaves the form of $M$ invariant\footnote{Though it does alter the interpretation of the components of $M$ so that velocities $\leftrightarrow$ electiric fields and rotations $\leftrightarrow$ magnetic fields under the action of T-duality.}.

For superstrings we include fermions so that 
\be
\Bd = |B\rangle_{bosonic} |B\rangle_{fermionic}
\ee
The bosonic conditions remain unchanged. The fermionic ones are given by
\be
\left ( \partial_{\sigma}\psi \pm \calM \partial_{\tau}\psi \right) \Bd = 0
\ee
For the non-zero modes the matrix $\calM_{\mu\nu}$ is the same as the bosonic case and the fermion sections of the boundary state are given by
\begin{eqnarray}
|B,\eta \rangle_{NSNS} = {\rm exp} \left\{i\eta \sum_{r>0} \psi^{\dagger\mu}_r \calM_{\mu\nu} \tilde{\psi}^{\dagger\nu}_r \right\} |0\rangle_{NSNS}	\\
|B, \eta \rangle_{RR} = {\rm exp} \left\{i\eta \sum_{r=0} \psi^{\dagger\mu}_r \calM_{\mu\nu} \tilde{\psi}^{\dagger\nu}_r \right\} |p, \eta\rangle_{RR}
\end{eqnarray}

The matrix $\calM$ has some useful properties: in the generic case with all possible Lorentz transformations applied and electro--magnetic fields turned on, it sits in the group $SO(1,9)$. However, for various combinations, these transformations and fields can be used to break it into a product of $SO(1,k)$ with other $SO$ subgroups, of both even and odd dimension. Effectively this means that we can rewrite $\calM$ in a block diagonal form based on this subgroup structure. In particular one can group all the space dimensions for which there is a boost or electric field into the $SO(1,k)$ subgroup, ignoring the action of the matrix $S_{\mu\nu}$ \cite{Callan:1996xx}.

A second property of $\calM$ is that the $SO(1,9)$ normalizing factor is $1/\det(\calG + \calF) \equiv 1/\bidet$, the inverse of square of the Born--Infeld action. If $\calM$ can be reduced to a block diagonal form, then $\bidet$ can be rewritten as $\pdet$, where $k$ labels the individual subgroups. This is useful for simplifying the handling the zero modes in the presence of fields and transformations.

\subsection{Zero Modes}
In the presence of a non--trivial matrix $\calM$, care needs to be taken with how it modifies the vacuum properties. The simplest way to approach this to define the generators of $\calM$:
\be
J^{\mu\nu} = l^{\mu\nu} + E^{\mu\nu} + K^{\mu\nu}\\
\ee
where
\begin{eqnarray}
l^{\mu\nu} &=& q^{\mu}p^{\nu}-q^{\nu}p^{\mu} \nonumber  \\
E^{\mu\nu} &=& -i \sum_{n=1}^\infty (\alpha^{\dagger\mu}_{n}\alpha^{\nu}_n -\alpha^{\dagger\nu}_{n}\alpha^{\mu}_n  +\tilde{\alpha}^{\dagger\mu}_{n}\tilde{\alpha}^{\nu}_n  -\tilde{\alpha}^{\dagger\nu}_{n}\tilde{\alpha}^{\mu}_n ) \nonumber  \\
K_{NSNS}^{\mu\nu} &=& -i \sum_{r > 0}^\infty (\psi^{\dagger\mu}_{r}\psi^{\nu}_r -\psi^{\dagger\nu}_{r}\psi^{\mu}_r  +\tilde{\psi}^{\dagger\mu}_{r}\tilde{\psi}^{\nu}_r  -\tilde{\psi}^{\dagger\nu}_{r}\tilde{\psi}^{\mu}_r ) \nonumber \\
K_{RR}^{\mu\nu} &=& {i \over 2}\left( [\psi_0^{\mu},\psi_0^{\nu}] + [\tilde{\psi}_0^{\mu},\tilde{\psi}_0^{\nu}]   \right)  
 -i \sum_{m= 1}^\infty (\psi^{\dagger\mu}_{m}\psi^{\nu}_m -\psi^{\dagger\nu}_{m}\psi^{\mu}_m  +\tilde{\psi}^{\dagger\mu}_{m}\tilde{\psi}^{\nu}_m  -\tilde{\psi}^{\dagger\nu}_{m}\tilde{\psi}^{\mu}_m ) \nonumber 
\end{eqnarray}
A finite transformation is then of the form
\begin{equation}
U(C) = \exp (iC_{\mu\nu}J^{\mu\nu})
\end{equation} 
which acts on the non--zero modes as
\be
\e^{i C \cdot J} a_n \cdot \tilde{a}_n \e^{-iC \cdot J} = a^{\mu}_n \calM_{\mu\nu}(C) \, \tilde{a}^{\nu}_n
\ee
{\it (a) Bosonic Zero Modes}.\newline
In ten dimensions there are no momentum or winding mode contributions to the bosonic zero mode. However, there are position states to be considered in the Dirichlet directions:
\be
\Bdz_X = \delta^\perp ({\bf q}) \Bz
\ee
Appling the generators we have
\be
\e^{i C \cdot J} {\bf q} = N{\bf q}
\ee
where $N^2 \equiv \calM$, so that
\be
\delta^\perp ({\bf q}) \rightarrow \delta^\perp (N{\bf q })
\ee
This can be easily shown to recover the pure Lorentz boost solution of \cite{Billo:1997eg} by noting that $q$ in the longnitudinal direction is identically zero\footnote{When considering the interaction of two D--branes at different positions ${\bf y,}$ one must also remember to Lorentz boost these: see ref \cite{Billo:1997eg, Kazama:1997bc} for details.}. This works independently of how we structure $\calM$. 

Since $\calM \propto 1/\bidet$ then it follows $N\propto 1/\rdet$. The properties of delta functions allow us to extract this to obtain the correct Born--Infeld term for the D--brane:
\be
\Bdz_X = \rdet \delta^\perp (\hat{N}{\bf q}) \Bz
\ee 
with $\langle 0 \Bz = 1$ \\
\newline
{\it (b) Fermionic Zero Modes}. \newline
The NSNS ground state is unaffected by the generators and we have the conventional normalization $_{NSNS}\langle 0 \Bz_{NSNS} = 1$. In the RR sector the ground states, $|B_\psi, \eta \rangle^{(0)}_{RR}$ are defined as
\be
|B_\psi, \eta \rangle^{(0)}_{RR} = \calM^{(\eta)}_{P \tilde{P}} | P = - {1 \over 2} \rangle | \tilde{P} = - {3 \over 2} \rangle 
\ee
with
\be
\calM^{(\eta)} = C \Gamma^0 \Gamma^{l_1} \ldots \Gamma^{l_p} \left( {1+ i\eta \Gamma_{11} \over 1+ i\eta} \right) 
\ee
where $C$ is the charge conjugation matrix and the $l_i$ lie in the space directions of the D$p$--brane world volume. The conjugate state is given by
\be
{}^{(0)}_{RR}\langle B_\psi, \eta| = \langle P = - {1 \over 2}| \langle \tilde{P} = - {3 \over 2}| \bar{\calM}^{(\eta)}_{P \tilde{P}}
\ee
such that
\begin{eqnarray}
\bar{\calM}^{(\eta)} &=& {\Gamma^0}^T \calM^{(\eta)} \Gamma^0 \nonumber \\
		     &=& (-1)^pC \Gamma^0 \Gamma^{l_1} \ldots \Gamma^{l_p} \left( {1- i\eta \Gamma_{11} \over 1+ i\eta} \right) 
\end{eqnarray}
When the fields and transformations are turned on $\calM^{(\eta)}$ becomes 
\be
C \Gamma^0 \Gamma^{l_1} \ldots \Gamma^{l_p} \prod_k { 1 - \sum v^k_{rs}\Gamma^r\Gamma^s \over \rkdet} \left( {1+ i\eta \Gamma_{11} \over 1+ i\eta} \right)
\ee
where
\be
v^k_{rs} = \kdet \times {1 \over 2} \calM^k_{rs} \;\; \mbox{for $r > s$}
\ee
If the subgroup is equivalent to the identity then we can ignore these insertions by setting $v_{rs}$ to zero.

\section{Amplitudes}
To obtain physical branes we need to impose the GSO projection \cite{Polchinski:1988tu}, which gives 
\be
|Dp\rangle =|Dp\rangle_{NSNS} +  |Dp\rangle_{RR}
\ee
such that\footnote{The relative normalization is verified by comparison with the open string amplitudes.}
\begin{eqnarray}
|Dp\rangle_{NSNS} &=& {1 \over 2} \left( |B,+ \rangle_{NSNS} - |B,- \rangle_{NSNS} \right)	\\
|Dp\rangle_{RR} &=& 2i  \left( |B,+ \rangle_{RR} + |B,- \rangle_{RR} \right) 
\end{eqnarray}

The tree level interaction between two D--branes with potentially different fields and boosts is given by 
\be
\calA = \langle Dp_2, M_2 |\Delta | Dp_1, M_1 \rangle = \calA_{NSNS} + \calA_{RR}
\ee
where
\begin{equation}
\Delta = {\alpha^\prime \over 4\pi} \int_{|z|\leq 1} {d^2z \over |z|^2} z^{L_0-a}\bar{z}^{\tilde{L}_0-a}
\end{equation}
is the closed string propagator used for calculating the amplitude $\langle B_2|\Delta |B_1\rangle$. $a=1$ for the bosonic string. For superstrings we have $a_{NS}=1/2$ and $a_R=0$. 

For our purposes we simply need 
\be
\calA \propto {\bf NS} - {\bf R}
\ee 
where
\begin{eqnarray}
{\bf NS} = &\prodn& {\det (1+q^{2n-1}\calM_2 \calM^T_1)(1-q^{2n})^2 \over \det (1-q^{2n}\calM_2 \calM^T_1)(1+q^{2n-1})^2 }  \nonumber \\
- &\prodn& {\det (1-q^{2n-1}\calM_2 \calM^T_1)(1-q^{2n})^2 \over \det (1-q^{2n}\calM_2 \calM^T_1)(1-q^{2n-1})^2 }
\end{eqnarray}
and
\begin{eqnarray}
{\bf R} = &\prodn& {\det (1+q^{2n}\calM_2 \calM^T_1)(1-q^{2n})^2  \over \det (1-q^{2n}\calM_2 \calM^T_1)(1+q^{2n-1})^2}  \nonumber \\
\mp &\prodn& {\det (1-q^{2n}\calM_2 \calM^T_1)(1-q^{2n})^2 \over \det (1-q^{2n}\calM_2 \calM^T_1)(1-q^{2n})^2}
\end{eqnarray}
The product terms not included in the determinants are the non--zero mode ghost and superghost contributions, while the $\mp$ sign depends on whether we are dealing with a brane - brane or brane - anti--brane configuration. 

Using the character properties of untwisted affine Lie algebras \cite{Fuchs:1992nq} this can be rewritten in terms of theta functions:
\begin{eqnarray}
{\bf NS} &=& f_1^2(q)\left( {1 \over f^2_3(q)} \prod_{j=1}^{5}{\Theta_{0,1}(z_j|q) \over \Theta_{1,1}(z_j|q)}-{1 \over f^2_4(q)} \prod_{j=1}^{5}{\Theta_{0,0}(z_j|q) \over \Theta_{1,1}(z_j|q)} \right) 	\\
{\bf R} &=&  f_1^2(q) \left( {1 \over f^2_2(q)}\prod_{j=1}^{5}{\Theta_{1,0}(z_j|q) \over \Theta_{1,1}(z_j|q)} \pm {1 \over f^2_1(q)} \prod_{j=1}^{5}{\Theta_{1,1}(z_j|q) \over \Theta_{1,1}(z_j|q)} \right)
\end{eqnarray}
where the $f_i(q)$ here are the ghost and superghost contributions; and the $\e^{2\pi z_j}$ are the parameters of $\calM_2 \calM^T_1$. For example, when dealing with a Lorentz boost then the corresponding $z_j$ is the rapidity. If $\calM_1 = \calM_2$ then, by the $SO(1,9)$ symmetry inherent in these, $\calM \calM^T = I$ and the standard results in terms of the $f_i$ functions are recovered. There is no need to distinguish between the NN, DD and ND modes since these are incorporated via the presence of $S$ in $\calM$.

As noted in \cite{Bertolini:1998mg, Billo:1998pg}, the inclusion of a non--zero odd Ramond spin structure in the amplitude requires not only pairing up of the co--ordinates but that at least one of the co--ordinates is properly inserted, say the 5th pair, {\it i.e.} there are no transformations or fields turned on for this pair of directions. This technical point is highly significant as it means that at least one of the set of $\Theta$ functions simplifies to cancel out the ghost and superghost modes. Similarly, when considering branes at angles to each other there are only four independant non--trivial angles to be considered, so again there is simplification which cancels the ghost modes. In all known examples of non--zero odd spin structure to date there seems to be such mechanisms at work to provide these cancellations. it is the presence of the ghost modes which stop the odd RR sector from disappearing altogether. In terms of $\mm$ this requires there to exist at least one trivial subgroup, $I_2$, of $SO(1,9)$. 

If a second $z_j$ is also zero then this is no longer true and the contribution vanishes. Nevertheless, in the approach taken in this paper this can be seen as the exception rather than the rule, so from now on we assume that all four of the $z_j$ are non-zero. The resulting amplitude is
\be
{\bf NS - R} = \prod_{j=1}^{4}{\Theta_{0,1}(z_j|q) \over \Theta_{1,1}(z_j|q)} - \prod_{j=1}^{4}{\Theta_{0,0}(z_j|q) \over \Theta_{1,1}(z_j|q)} - \prod_{j=1}^{4}{\Theta_{1,0}(z_j|q) \over \Theta_{1,1}(z_j|q)} \pm  \prod_{j=1}^{4}{\Theta_{1,1}(z_j|q) \over \Theta_{1,1}(z_j|q)}
\ee

This is much more useful as the majority of known Theta function identities are of order 4. 

Before we examine the supersymmetry of these configurations we need to deal with several issues arising out of the contribution to the amplitude of the RR zero modes. 
\subsection{Regularization of Fermion Zero Modes} 
So far we have implicitly allowed two D--branes of arbitary dimensions to interact. However, a naive examination of the normalization of the RR zero modes shows that the only configurations permitted are where the number of mixed ND boundary conditions, $\nu$, are equal to 0 or 8 \cite{Bergman:1997gf}. A more sophisticated approach to this problems requires the use of the regularization technique introduced in \cite{yost} and developed for D--branes in \cite{Billo:1998vr}, where RR zero modes and superghost modes are dealt with together. This technique has been used to show the existence of the important $\nu = 8$ solutions and also $\nu = 6$ solutions \cite{Billo:1998ew}. To get a more general set of solutions is not trivial as the regularization technique requires the appropriate pairing up of dimensions. In the literature to date this pairing has been intuitive, based on the natural $SO(2)$ subgroup formation present in the matrices $\calM$ being chosen. In these $\calM$ is already diagonal and as a result there is no mixing of N and D directions to break the $SO(2)$ substructures. These $SO(2)$ subgroups are different in origin from the ones that are required to cancel the ghost modes and in the generic configuration it is not possible to identify them.

If $\mm$ has no natural $SO(2)$ substructures, then the intuitive argument breaks down as the pairing can no longer simply be mapped from $\mm$ to the spacetime. The resolution to problem is to choose a pairing based on the affine $SO(1,9)$ parameters $z_j$ appearing in the non--zero mode contributions to the amplitude. That is, choose a spacetime basis whereby each $z_j$ also labels a pair of dimensions ${\bf a}_j = (a_{j_1}, a_{j_2})$. It is this pairing that the fermion number operators, $N_j = \Gamma^{a_{j_1}}\Gamma^{a_{j_2}}$ are now constructed as outlined in \cite{Billo:1998vr}. In this way the overall structure of the configuration, encoded in $\mm$, is utilized as opposed to the structure of the individual branes. Hence
\be
{\bf F_0} = {1 \over 2} \sum_{k=1}^5 N_k
\ee
The regularizing factor used is ${\cal R}(x) = x^{2({\bf F_0 + G_0})}$ in the limit $x \rightarrow 1$; ${\bf F_0}$ and ${\bf G_0}$ are the fermion and superghost number operators respectively. Thus we have 
\be
 {}^{(0)}_{RR}\langle B^1, \eta_1  |B^2, \eta_2 \rangle^{(0)}_{RR} 
= \lim_{x \rightarrow 1} \: {}^{(0)}_{RR}\langle B^1, \eta_1 | {\cal R}(x)|B^2, \eta_2 \rangle^{(0)}_{RR}
\ee
with 
\be
|B^2, \eta_2 \rangle^{(0)}_{RR} = |B^2_{\psi}, \eta_2 \rangle^{(0)}_{RR} \: |B^2_{sgh}, \eta_2 \rangle^{(0)}_{RR}
\ee
The superghost expression is independent of $\mm$, {\it viz. }
\begin{eqnarray}
{}^{(0)}_{RR}\langle B_{sgh}^1, \eta_1 |\Delta |B^2_{sgh}, \eta_2 \rangle^{(0)}_{RR} &=& \langle -{3 \over 2}, -{1 \over 2} | \e^{i\eta_1\beta_0 \bar{\gamma}_0} x^{-2\gamma_0\beta_0} \e^{i\eta_2\gamma_0\bar{\beta}_0} |-{1 \over 2}, -{3 \over 2}\rangle \nonumber \\
&=& {1 \over 1-\eta_1\eta_2x^2}
\end{eqnarray}
The RR fermion zero mode amplitude becomes
\be
{}^{(0)}_{RR}\langle B_{\psi}^1, \eta_1 |\Delta |B^2_{\psi}, \eta_2 \rangle^{(0)}_{RR} = \mbox{tr} \left( x^{2{\bf F_0}}  \calM^{(\eta_1)} C^{-1} \calM^{(\eta_2)^T} C^{-1} \right)
\ee
Substituting in for the $\calM^{(\eta)}$ explicitly, this becomes
\begin{eqnarray}
-{1 \over \sqrt{\det(C_1)} \sqrt{\det(C_2)}}\mbox{tr}  \left\{ x^{2{\bf F_0}} \left( 
\prod_\alpha \Gamma^\alpha \delta_{\eta_1\eta_2, -1} + (-1)^p \prod_\alpha \Gamma^\alpha \Gamma^{11}  \delta_{\eta_1\eta_2, +1} \right) \times \right. \nonumber \\
\left. \left( (1+ \sum_{r>s} v^1_{rs}\Gamma^r\Gamma^s + \sum_{t>u} v^2_{tu}\Gamma^t\Gamma^u+(\sum_{r>s} v^1_{rs}\Gamma^r\Gamma^s)(\sum_{t>u} v^2_{tu}\Gamma^t\Gamma^u) \right)  \right\}
\end{eqnarray}
where $\alpha$ labels the $\Gamma$ matrices that correspond to the $\nu$ ND mixed directions. The normalizing term, $1 / \sqrt{\det(C_1 C_2)}$, cancels the bosonic term normalization as desired. Expanding this out we see that, depending on the fields and transformations present, there is a variety of combinations of $\Gamma$s remaining, denoted by $\beta$, giving rise to individual expressions of the form
\begin{eqnarray}
\mbox{tr} \left( x^{2{\bf F_0}}\prod_\beta \Gamma^\beta \right) &=&  \prod_{k=1}^{{10-\rho \over 2}} \mbox{tr}\left( x^{N_k}  \right) \prod_{l=1}^{{\rho \over 2}} \mbox{tr} \left( x^{N_l}N_l  \right) \nonumber \\
&=& \left( x+ {1 \over x} \right)^{{10-\rho \over 2}} \left(x- {1 \over x} \right)^{{\rho \over 2}} 
\end{eqnarray}
and
\begin{eqnarray}
\mbox{tr} \left (x^{2{\bf F_0}}\prod_\beta \Gamma^\beta \Gamma^{11} \right) &=& \prod_{l=1}^{{\rho \over 2}} \mbox{tr}\left( x^{N_l}  \right) \prod_{k=1}^{{10-\rho \over 2}} \mbox{tr} \left( x^{N_k}N_k  \right) \nonumber \\
&=& \left( x+ {1 \over x} \right)^{{\rho \over 2}} \left(x- {1 \over x} \right)^{{10-\rho \over 2}}
\end{eqnarray}
where $\rho$ counts the number of $\Gamma^\beta$ and we have used $\Gamma^{11} = \prod_{i=j}^{5} N_j$.

Combining these with the contribution from the superghosts and taking the limit $x \rightarrow 1$ we have
\begin{eqnarray}
{}^{(0)}_{RR}\langle B^1, \eta_1  |B^2, \eta_2 \rangle^{(0)}_{RR} 
&\propto& - \lim_{x \rightarrow 1} \:
 \left[  \left(  x+ {1 \over x} \right)^{{10-\rho \over 2}} \left( x- {1 \over x} \right)^{{\rho \over 2}} {1 \over 1+x^2}\delta_{\eta_1\eta_2, -1} \right.
\nonumber \\
& & \;\;\;\;\;\;\;\;\;\;\;\;\;+\left.  \left(  x+ {1 \over x} \right)^{{\rho \over 2}} \left(x- {1 \over x} \right)^{{10-\rho \over 2}}{1 \over 1-x^2}\delta_{\eta_1\eta_2, +1} \right] \nonumber \\
&\propto& - 16\delta_{\rho,0}\delta_{\eta_1\eta_2, -1} +  16\delta_{\rho,8}\delta_{\eta_1\eta_2, +1}
\end{eqnarray}
Since there are always multiples of two $\Gamma$s in the expression, $\rho$ is always even, keeping in line with the D--branes allowed under the GSO projection\footnote{The $+$ sign is symbolic as the two expressions will have different constants coming from $\mm$ and there are potentially many such expressions in the amplitude.}. The expression looks similar to the one derived for a static brane with all fields turned off, but it is important that $\rho$ is not confused with $\nu$ though in some cases it is possible to identify the two. 

More significant in these calculations is whether the fields/transformations are applied or not, as opposed to their actual values. As there are contributions of $\Gamma$s in sets of two and four, one can have $\nu = 0, 2, 4$ for brane - brane systems and $\nu =  8, 6, 4$ for brane - anti--brane systems. Only for $\nu = 4$ does it seem possible to have both brane - brane and brane - anti--brane systems. It was noted in the appendix of \cite{Bergman:1997gf} that one can perform a rotation of a brane through $\pi$ in a plane so that it becomes an anti--brane. This rotation is however a Lorentz transformation, so is a specific case of above and the change in $\calM^{(\eta)}$ needs to be taken into account.

\section{Supersymmetry}
Order 4 identities for Theta functions come from the so called Riemann's Theta Identities. The generic derivation for these starts with an integer valued matrix, $A$, satisfying
\be
A^TA = m^2 I_n\,; \:\: m, n \in {\bf Z} \label{eq:riemann}
\ee
Applied to Theta functions we obtain the general rule
\be
m \prod_{i=1}^n \theta(v_i) =  \sum_\zeta \kappa_\zeta \left( \prod_{i=1}^n \theta (u_i + \zeta) \right)
\ee
where
\begin{eqnarray}
\zeta &=& 0, {1 \over 2}, {1 \over 2}\tau, {1 \over 2}(1+\tau) \nonumber \\
\kappa_\zeta &=& \e^{i\pi (\tau + \sum_i u_i)} \\
v_i &=& A u_i \nonumber
\end{eqnarray}
Substituting in for $\zeta$ directly we recover by definition $\Theta_{a, b}, \: a,b = 0,1$ so it is clear that we can map the expression for ${\bf NS} - {\bf R}$ to an appropriate matrix $A$. As there are plus and minus signs in ${\bf NS - R}$ we must first perform modular transformations and multiply by appropriate terms to obtain an exact correlation. In the above case this has already been performed and is given as equation $R_5$ on page 18 of \cite{mumford}; {\it i.e.}, replace one of the $u_i$ by $u_i+\tau+1$ and multiply by $\exp(i\pi\tau + 2i\pi u_i)$. The final result for a brane -- brane configuration, is 
\be
{\bf NS - R} = \prod_{i=1}^n \Theta_{1,1}(v_i)
\ee 
The importance of this is that $ \Theta_{1,1}(0) =0$, and this is the only $\Theta_{a,b}$ for which this is true. Returning to the  equation defining the identity, the set of solutions giving this is determined by putting each one of the $v_i = A_{ij}u_j$ equal to zero.

When the amplitude vanishes a fraction of the supersymmetry is preserved. The corollary is that when it is not satisfied we are dealing with a system with completely broken supersymmetry. We also have to take into account that there are special values of the $z_j$ which can also give non-zero supersymmetric solutions. 

For a given system of two interacting branes, if one of the defining equations for $v_i$ holds such that
\be
\sum_{i=1}^4 v_i z_i = 0 \label{eq:susysum}
\ee 
then the fraction of supersymmetry preserved is $1/16$ as a Killing spinor can survive \cite{SheikhJabbari:1997cv}. The rule is as follows: if the $z_i$ are such that $k$ equations of the form $\sum v_i z_i = 0$ are satisfied then the fraction of supersymmetry preserved is $k/16$. This allows us to recover preserved supersymmetries up to $1/4$. It is not possible to identify each fraction preserved with a particular interaction between the branes, as the vanishing of the sum is over all parameters $z_j$.

 To get higher fractions requires the special case setting all $z_i = 0$, which preserves $1/2$ of the supersymmetries. This corresponds to when the branes are parallel, move with the same velocity and have identical fields so that the BPS condition, $\mm = I$, is recovered. 

The solution given in \cite{SheikhJabbari:1997cv} corresponds to the matrix $A$ used in deriving the original Riemann Theta Identities which can be found in \cite{mumford}. Explicitly: 
\be
A = 
\left( 
\begin{array}{crrr}
1 & 1 & 1 & 1 \\
1 & 1 & -1 & -1 \\
1 & -1 & 1 & -1 \\
1 & -1 & -1 & 1    
\end{array} 
\right)
\ee
which translates into the supersymmetry preserving conditions 
\begin{eqnarray}
z_a+z_b+z_c+z_d &=& 0 \nonumber \\
z_a+z_b-z_c-z_d &=& 0 \nonumber \\
z_a-z_b+z_c-z_d &=& 0 \\
z_a-z_b-z_c+z_d &=& 0  \nonumber
\end{eqnarray}
These have been classified according to the amount of supersymmetry they preserve in \cite{SheikhJabbari:1997cv} in terms of rotated brane angles\footnote{note that these solutions are all mod $\pi$.}, however utilizing T-duality they can equivalently considered as turning on electric--magnetic fields. 

A complete classification of supersymmetries for these type of configurations is thus obtained by determining all the possible $A$. Given a set of integers $\{ v_i \} $ for which~(\ref{eq:susysum}) holds and also satisfy 
\be
\sum_{i=1}^4 v_i^2 = m^2
\ee 
as follows from~(\ref{eq:riemann}), the existance of $\{ v_i \}$ is a necessary condition for the preservation of some supersymmetry. For a sufficient condition one needs to construct the full integer matrix $A$ or at least show that it exists, though~(\ref{eq:susysum}) need not hold for all rows.

If no supersymmetry is preserved then we can still use the existence of the Riemann Theta Identities to simplify ${\bf NS - R}$ to
\be
 m \prod_{i=1}^4 \left[ {\Theta_{1,1}((Az)_i) \over \Theta_{1,1}(z_i)} \right]
\ee

For the brane - anti--brane case the above analysis does not hold as there is no substitution $z_j \rightarrow z_j + a\tau + b$ which will give the correct set of signs. However, if we rewrite, symbolically, the amplitude as
\be
\left[ NS(+) + NS(-) + R(+) + R(-) \right] - 2R(-)
\ee
and if some supersymmetry is preserved then the term in square brackets vanishes and we are left with an amplitude proportional to 
\be
\prod_{i=1}^4\left[ {\Theta_{1,1}(z_i) \over \Theta_{1,1}(z_i)} \right]
\ee
which cancels to unity. This implies that there is the possibility for the creation of a fundamental string in these configurations as the branes cross through each other adiabatically \cite{Kitao:1998vn, Bergman:1997gf} as long as some supersymmetry holds. The created string will have properties as defined by the $z_j$.

\section{Conclusion}
 In this paper we have shown how Riemann Theta Identies can be used to classify the preserved supersymmetry of arbitary configurations of D-branes with non--vanishing odd spin structure. This has been done by using the identities to map the amplitudes to integer matrices $A$. Any supersymmetric configuration must be such that a matrix $A$ can be constructed which satisfies a simple but non--trivial equation. As the amplitudes used are invariant under T--duality the results are quite general and can be applied to a variety of different configurations since it is the form of the amplitude that is significant, not how it was constructed. 

We have also derived a general rule for when the configurations are permitted to have a RR sector, which is intimately connected to the relative configuration of fields and transformations between the interacting branes encoded in $\mm$. Again this constraint is related to the overall configuration between the branes and not to them individually. This constraint will also result in restrictions on the types of bound states permitted.

Though the cancellation of ghosts and superghosts allow solutions to be constructed, the physical understanding of this is non--trivial. For instance, D0--branes would expect to have five non-zero $z_j$, permitted. The co--ordinate pair that cancels the ghosts do not have to be in the light cone directions, which would imply that this feature is independant of whether we use light cone or covariant quantization, {\it i.e.} the fact the D0--brane is defined in term of a superstring is not significant in this case. What controls the presence of supersymmetry is the subgroup properties of $\mm$ which is still required to have a trivial $SO(2)$ subgroup for any supersymmetry to be preserved at all. Moreover, the subgroup properties are generically independant of the subgroup properties of the constituent matrices $\calM$. However, it is not obvious how this can generically be related directly the rules governing which $p - p^\prime$ configurations are permitted, if at all. 

Interestingly there does not appear to exist any $5\times 5$ matrix satisfying the conditions for $A$, and thus no corresponding order 5 Riemann Theta Identities which would point towards supersymmetry preserving solutions for five non--zero $z_j$. 
\newline
{\bf Acknowledgements} Thanks to David Dunbar, Stephen Howes and Karl Lloyd for discussions during the course of this paper. This work was supported by PPARC.

\end{document}